\documentclass{article}

\input{tcilatex}
\begin{document}

\title{A Free-Field Lagrangian for a Gauge Theory of the CPT Symmetry}
\author{Kurt Koltko\\localcpt@yahoo.com}
\date{March 10, 2017\\revised April 25, 2017\\revised July 19, 2018}
\maketitle

\begin{abstract} A simplified mathematical approach is presented and used to find a suitable
free-field Lagrangian to complete previous work on constructing a gauge
theory of CPT transformations. \ The new Lagrangian is a slight but
important modification of the previous candidate. \ The new version
satisfies an additional requirement which had been overlooked and not
satisfied by the previous candidate. \end{abstract}
\bigskip

INTRODUCTION
\bigskip

Aside from curiosity, why consider gauging the CPT symmetry? \ We assert
that gauging the CPT symmetry is necessary to further our understanding of
gravitational physics. \ The fact that the metric spin connection
formulation of general relativity can be derived from gauging the global,
proper, continuous Lorentz transformations ($\Lambda $) should be sufficient
motivation to consider gauging CPT. \ This is because PT is also a proper
Lorentz transformation, and it would seem logical to include PT in gauging
the \textit{entire }proper Lorentz group of transformations, $PT\Lambda $. \
It is necessary to include the C operation because PT is not a universal
symmetry, whereas CPT \textit{is} a universal symmetry. \ In effect, we are
gauging the $CPT\Lambda $ transformation of the Dirac field to induce the
gauging of the full group of proper spacetime Lorentz transformations.

Another reason to gauge the CPT symmetry is because some gravitational
issues can be interpreted as requiring an additional force, and the gauging
of experimentally verified, global symmetries has been successful in
determining the known forces. \ Given the historical success from gauging
symmetries, we do not have much (if any) choice if one is to continue down
that path because CPT is a universal, experimentally verified symmetry that
does not require any extra dimensions.

For example, we argue that gauge CPT is a logical alternative to the
galactic dark matter hypothesis used in the context of spiral galaxies [1,
2]. \ The dark matter is invoked to explain galactic rotation curves for two
reasons, really. \ The primary reason is that the gravity produced by the
observed galactic matter does not account for the motion of material (stars,
HI gas) as the distance from the galactic center increases. \ However,
another reason is buried within - a mass independent acceleration. \ This
acceleration is what steers a hypothetical explanation towards dark matter
rather than a missing force - no forces other than gravity produce a mass
independent acceleration. \ Thus, if gravity is to explain this strange
motion, then there must be missing, unseen matter. \ However, another
possible explanation could logically exist in the form of a missing force or
an extension of general relativity because, after all, accelerations are
produced by forces.

The traits of universality and of containing a proper Lorentz transformation
that CPT share with $\Lambda $ are the heuristic motivation that a mass
independent acceleration (i.e., the equivalence principle is obeyed) would
occur if any new forces or extensions of general relativity were to be
unveiled by gauging $CPT\Lambda $. \ Indeed, the new gauge field $X_{\mu }$
introduced to accommodate local $CPT\Lambda $ transformations is required to
contain terms of the form $x_{\mu ab}\sigma ^{ab}$ ($\sigma ^{ab}=\frac{1}{4}%
\left[ \gamma ^{a},\gamma ^{b}\right] $) which appear along with the metric
spin connection containing term, $\omega _{\mu ab}\sigma ^{ab}$, of general
relativity.\footnote{Greek indices such as $\mu $ denote manifold
coordinates. \ Latin indices such as $a$ denote the local inertial
coordinates. \ Otherwise, we use the Bjorken-Drell conventions except for $%
\sigma ^{ab}$.}  Hence, the spinor fields $\psi $ describing
matter will interpret $x_{\mu ab}\sigma ^{ab}$ as a gravitational effect. \
More detailed arguments regarding gauge CPT as an alternative to the dark
matter hypothesis can be found in [1, 2].

Furthermore, gauging CPT should be of interest in the broader problem of
reconciling general relativity with quantum theory. \ That the CPT symmetry
is born neither from special relativity nor quantum theory alone, but rather
the phenomenologically successful union of the two, warrants paying closer
attention to CPT. \ Because general relativity can be obtained by making
special relativity (i.e. $\Lambda $) local, it would seem interesting to
include CPT when making $\Lambda $ local in order to incorporate quantum
theory at a fundamental level. \ This is analogous to expanding the early
nonrenormalizable weak interaction theories by the renormalizable $SU\left(
2\right) \times U\left( 1\right) $ electroweak theory - any approach to
quantum gravity would need to include the missing spacetime dynamical
degrees of freedom unveiled by gauging CPT.
\bigskip

THE GAUGE THEORY OF CPT TRANSFORMATIONS
\bigskip

Gauging $CPT\Lambda $ is a straightforward process analogous to gauging
other symmetries. \ The original approach ([1, 2]) used variational
principles in conjunction with the action integral primarily because of the
appearance of Dirac delta functionals when making the discrete, global CPT
symmetry a local symmetry transformation. \ Dirac delta functionals only
have meaning when appearing in a definite integral, and the action integral
is the only obvious, natural arena for the Dirac delta functionals which
occur. \ Also, we view the action integral to be \textit{the} object of
fundamental importance when dealing with symmetries. \ For a more in-depth
discussion of motivation, experimental possibilities, and derivation of the
mathematics used to handle these discrete transformations, see [1, 2].

We begin with the global $CPT\Lambda $ transformation applied to the Dirac
spinor $\psi $ and the spacetime vierbein $e_{a}^{\;\mu }$: \ $\psi
\rightarrow i\gamma ^{5}\Lambda _{\psi }\psi $, and $e_{a}^{\;\mu
}\rightarrow -e_{b}^{\;\mu }\Lambda _{a}^{\;b}$, where $\Lambda _{\psi }$
and $\Lambda _{a}^{\;b}$ are the spinor and spacetime representations of $%
\Lambda $. \ The $CPT$ transformation appears as the $i\gamma ^{5}$ in the
spinor transformation and as the factor of $-1$ (due to $PT$) appearing in
the vierbein transformation. \ We assume there is no non-trivial \textit{%
spacetime} analog of the C operation, i.e., there is no such thing as
"anti-spacetime" (an attempt to find a non-trivial spacetime C operation is
contained in [2]). \ There are two steps in making these transformations
local. \ First, we make $\Lambda $ a local transformation. \ Then, we
introduce a real, differentiable function $f$ \ to be used as the argument
of unit step functions $\Theta $, where we define $\Theta \left[ f\leq 0%
\right] =0$, and $\Theta \left[ f>0\right] =1$. \ In \textit{arbitrarily
chosen} regions where we want to perform the $CPT\Lambda $ operation, we set 
$f>0$. \ In the regions we leave alone, we set $f<0$. \ The boundaries
between the regions where $CPT\Lambda $ is applied and where it is not are
given by $f=0$. \ It is to be emphasized that $f$ is \textit{not} a physical
field, just a parameter used to make CPT local; therefore, the function $f$
must disappear from the Lagrangian and ensuing field equations. \ One thus
obtains the local $CPT\Lambda $ transformations [1]:

\begin{eqnarray*}
e_{a}^{\;\mu } &\rightarrow &\Theta \left[ -f\right] e_{a}^{\;\mu }-\Theta %
\left[ f\right] e_{b}^{\;\mu }\Lambda _{a}^{\;b}, \\
\psi &\rightarrow &\Theta \left[ -f\right] \psi +\Theta \left[ f\right]
i\gamma ^{5}\Lambda _{\psi }\psi \equiv U\psi ,\text{ where} \\
U &=&\Theta \left[ -f\right] I+\Theta \left[ f\right] i\gamma ^{5}\Lambda
_{\psi }\text{.}
\end{eqnarray*}%
We denote the local $CPT\Lambda $ transformation by the customary $U$ even
though the transformation is not unitary. \ Unitarity is not necessary in
making gauge theories. \ For example, $\Lambda $ is not unitary and is used
to derive the metric spin connection formulation of general relativity as a
gauge theory. \ We also note that the presence or absence of a conservation
law associated with a global symmetry transformation has no relevance to
constructing the ensuing gauge theory. \ Again, returning to $\Lambda $, we
note that there are no conservation laws associated with global boosts.

To proceed further, we must take derivatives and make products from the
above transformations. \ Because of the discrete nature of the
transformations, various discontinuities naturally arise. \ The action
integral, variational methods, and elementary functional analysis lead to
the following important rules needed to handle the discontinuities ([1, 2]): 
$\Theta \left[ -f\right] \Theta \left[ f\right] =0$, $\left( \Theta \left[
\pm f\right] \right) ^{n}=\Theta \left[ \pm f\right] $ ($n$ is a positive
integer $\geq 1$), and $\partial _{\mu }\Theta \left[ \pm f\right] =\pm
\delta \left[ f\right] \partial _{\mu }f$, where $\delta \left[ f\right] $
is a Dirac delta functional. \ Application of these rules to the above
transformations immediately gives us:

\begin{eqnarray*}
U^{-1} &=&\Theta \left[ -f\right] I-\Theta \left[ f\right] i\gamma
^{5}\Lambda _{\overline{\psi }}, \\
\partial _{\mu }U &=&\Theta \left[ f\right] i\gamma ^{5}\partial _{\mu
}\Lambda _{\psi }+\delta \left[ f\right] \partial _{\mu }f\left( i\gamma
^{5}\Lambda _{\psi }-I\right) ,\text{ and} \\
\left( \partial _{\mu }U\right) U^{-1} &=&\Theta \left[ f\right] \left(
\partial _{\mu }\Lambda _{\psi }\right) \Lambda _{\overline{\psi }}+\Theta %
\left[ -f\right] \delta \left[ f\right] \partial _{\mu }f\left( i\gamma
^{5}\Lambda _{\psi }-I\right) \\
&&+\Theta \left[ f\right] \delta \left[ f\right] \partial _{\mu }f\left(
I+i\gamma ^{5}\Lambda _{\overline{\psi }}\right) ,
\end{eqnarray*}%
where $\Lambda _{\overline{\psi }}$ is the inverse of $\Lambda _{\psi }$
(also, $\overline{\psi }\rightarrow i\overline{\psi }\gamma ^{5}\Lambda _{%
\overline{\psi }}$ under global $CPT\Lambda $).

Everything follows from these equations. \ The Dirac equation is obviously
not invariant under $U$, so we introduce a minimally coupled gauge field as
part of a covariant derivative acting on $\psi $. \ As always, a gauge field 
$Z_{\mu }$ associated with $U$ must transform as $Z_{\mu }\rightarrow
UZ_{\mu }U^{-1}-\frac{1}{\beta }\left( \partial _{\mu }U\right) U^{-1}$,
where $\beta $ is the coupling constant. \ Because $\left( \partial _{\mu
}U\right) U^{-1}$ has terms containing $I$ and $\gamma ^{5}$in addition to $%
\sigma ^{ab}$, we immediately see from the linear independence of $I$, $%
\gamma ^{5}$, and $\sigma ^{ab}$ that the metric spin connection term, $%
\omega _{\mu ab}\sigma ^{ab}$, cannot compensate for local $CPT\Lambda $
transformations (see [1] for an alternative proof). \ Therefore, we
introduce a new gauge field $X_{\mu }$ of the form $X_{\mu }=x_{\mu
I}I+x_{\mu 5}\gamma ^{5}+x_{\mu ab}\sigma ^{ab}$ as part of a covariant
derivative $D_{\mu }=\partial _{\mu }+\frac{1}{2}\omega _{\mu ab}\sigma
^{ab}+\beta X_{\mu }$. \ In other words, $Z_{\mu }=\frac{1}{2\beta }\omega
_{\mu ab}\sigma ^{ab}+X_{\mu }$. \ One can easily verify that the following
transformation (see [1] for the derivation) for $X_{\mu }$ gives $D_{\mu
}\rightarrow UD_{\mu }U^{-1}$:

\begin{eqnarray*}
X_{\mu } &\rightarrow &\Theta \left[ -f\right] X_{\mu }+\Theta \left[ f%
\right] \Lambda _{\psi }X_{\mu }\Lambda _{\overline{\psi }}+\Theta \left[ -f%
\right] \delta \left[ f\right] Y_{\mu }+\Theta \left[ f\right] \delta \left[
f\right] \widetilde{Y}_{\mu },\text{ where} \\
Y_{\mu } &=&\beta ^{-1}\left[ \partial _{\mu }f\left( I-i\gamma ^{5}\Lambda
_{\psi }\right) -\frac{1}{2}\varsigma _{\mu ab}\sigma ^{ab}\right] , \\
\widetilde{Y}_{\mu } &=&\beta ^{-1}\left[ \partial _{\mu }f\left( -I-i\gamma
^{5}\Lambda _{\overline{\psi }}\right) -\frac{1}{2}\widetilde{\varsigma }%
_{\mu ab}\sigma ^{ab}\right] .
\end{eqnarray*}%
The terms $\varsigma _{\mu ab}$, $\widetilde{\varsigma }_{\mu ab}$ come from
the differentiation of the transformed vierbein contained in the metric spin
connection, $\omega _{\mu ab}$, under local $CPT\Lambda $:%
\[
\omega _{\mu ab}\rightarrow \Theta \left[ -f\right] \omega _{\mu ab}+\Theta %
\left[ f\right] \widetilde{\omega }_{\mu ab}+\Theta \left[ -f\right] \delta %
\left[ f\right] \varsigma _{\mu ab}+\Theta \left[ f\right] \delta \left[ f%
\right] \widetilde{\varsigma }_{\mu ab}\text{,} 
\]%
where $\widetilde{\omega }_{\mu ab}$ satisfies $\widetilde{\omega }_{\mu
ab}\sigma ^{ab}=\Lambda _{\psi }\omega _{\mu ab}\sigma ^{ab}\Lambda _{%
\overline{\psi }}-2\left( \partial _{\mu }\Lambda _{\psi }\right) \Lambda _{%
\overline{\psi }}$. \ Explicit expressions for $\omega _{\mu ab}$, $%
\widetilde{\omega }_{\mu ab}$, $\varsigma _{\mu ab}$, and $\widetilde{%
\varsigma }_{\mu ab}$ can be found in [1].

We are now ready to find the Lagrangian using the usual machinery of gauge
theories. \ We begin by deriving the candidate found in the previous
versions of this paper\footnote{This Lagrangian was first proposed in [1] as
the "peel-off" Lagrangian. \ However, gauge covariance was not
shown.} and then introduce the modification leading to the new
candidate. \ First, we know that $X_{\mu }$ must be massless (see [1] for an
alternative proof). \ We also know that $\left[ D_{\mu },D_{\nu }\right] $
transforms gauge covariantly. \ Unfortunately, using $Tr\left\{ \left[
D_{\mu },D_{\nu }\right] \left[ D^{\mu },D^{\nu }\right] ^{\dag }\right\} $
as a free-field term introduces $R^{2}$ into the Lagrangian instead of the
required $R$ of general relativity, where $R$ is the Einstein-Hilbert scalar
curvature term formed from just the metric spin connection and a couple of
vierbeins. \ Use of just $\left[ D_{\mu },D_{\nu }\right] $ is problematic
because it is not clear what to contract it with, courtesy of the absence of
the Latin (i.e. inertial) indices in the $x_{\mu I}$ and $x_{\mu 5}$ terms.
\ Also, at some point in the total Lagrangian, the $\omega _{\mu ab}$ and $%
x_{\mu ab}$ terms must appear on an unequal footing outside of the $\frac{1}{%
2}\omega _{\mu ab}\sigma ^{ab}+\beta x_{\mu ab}\sigma ^{ab}$ combination -
otherwise, the sole purpose of $x_{\mu ab}$ is just to be a "fudge factor"
introduced to get rid of the $\varsigma _{\mu ab}$, $\widetilde{\varsigma }%
_{\mu ab}$ terms appearing in the transformation of $\omega _{\mu ab}$.

The key to constructing the previous Lagrangian was that $R$ is invariant
[1] and covariant (shown below) under local $CPT\Lambda $ transformations
even though $\omega _{\mu ab}$ is not. \ (Briefly [1], the $R$ formed from
the metric spin connection, $\omega _{\mu ab}$, is the same as the familiar $%
R$ formed from the metric tensor $g^{\mu \nu }$, and the $R$ from $g^{\mu
\nu }$ is gauge invariant. \ Gauge invariance follows because $g^{\mu \nu
}=\eta ^{ab}e_{a}^{\;\mu }e_{b}^{\;\nu }\rightarrow \Theta \left[ -f\right]
g^{\mu \nu }+\Theta \left[ f\right] g^{\mu \nu }$, where $\eta ^{ab}$ is the
Minkowski metric tensor. \ From this we immediately see that derivatives of
the metric tensor also transform in the same manner, $\partial _{\rho
}g^{\mu \nu }\rightarrow \Theta \left[ -f\right] \partial _{\rho }g^{\mu \nu
}+\Theta \left[ f\right] \partial _{\rho }g^{\mu \nu }$. \ So, the metric
tensor and $R$ and any functions of these only pick up removable
singularities under local $CPT\Lambda $ transformations. \ These
singularities have no effect on the action integral and are therefore
dropped. \ Again, the action integral and variational approach ([1, 2]) are
of fundamental importance in dealing with singularities.)

We now show that the Einstein-Hilbert $R$ transforms gauge covariantly. \
Or, equivalently, we need to show that $R_{\mu \nu ab}\sigma
^{ab}\rightarrow UR_{\mu \nu ab}\sigma ^{ab}U^{-1}$ under local $CPT\Lambda $
transformations ( the transformation of the vierbeins used in the
contraction of $R_{\mu \nu ab}$ to obtain $R$ cause no complications because
they do not introduce any delta functionals or gamma matrices). \ We begin
with re-expressing $R_{\mu \nu ab}$ in terms of the more familiar form of
the Riemann curvature tensor: \ $R_{\mu \nu ab}=R_{\mu \nu \rho \sigma
}e_{a}^{\;\rho }e_{b}^{\;\sigma }$. \ We recall that $R_{\mu \nu \rho \sigma
}$ is invariant under local $CPT\Lambda $ transformations because $R_{\mu
\nu \rho \sigma }$ is comprised of the metric tensor and its derivatives. \
So, we have:

$R_{\mu \nu ab}\sigma ^{ab}=$ $R_{\mu \nu \rho \sigma }e_{a}^{\;\rho
}e_{b}^{\;\sigma }\sigma ^{ab}$

$\rightarrow R_{\mu \nu \rho \sigma }\left( \Theta \left[ -f\right]
e_{a}^{\;\rho }-\Theta \left[ f\right] e_{c}^{\;\rho }\Lambda
_{a}^{\;c}\right) \left( \Theta \left[ -f\right] e_{b}^{\;\sigma }-\Theta %
\left[ f\right] e_{d}^{\;\sigma }\Lambda _{b}^{\;d}\right) \sigma ^{ab}$

$=$ $R_{\mu \nu \rho \sigma }\left( \Theta \left[ -f\right] e_{a}^{\;\rho
}e_{b}^{\;\sigma }+\Theta \left[ f\right] \Lambda _{a}^{\;c}\Lambda
_{b}^{\;d}e_{c}^{\;\rho }e_{d}^{\;\sigma }\right) \sigma ^{ab}$.

We compare this with $UR_{\mu \nu ab}\sigma ^{ab}U^{-1}$:

$UR_{\mu \nu ab}\sigma ^{ab}U^{-1}$

$=\left( \Theta \left[ -f\right] I+\Theta \left[ f\right] i\gamma
^{5}\Lambda _{\psi }\right) R_{\mu \nu ab}\sigma ^{ab}\left( \Theta \left[ -f%
\right] I-\Theta \left[ f\right] i\gamma ^{5}\Lambda _{\overline{\psi }%
}\right) $

$=\left( \Theta \left[ -f\right] R_{\mu \nu ab}\sigma ^{ab}+\Theta \left[ f%
\right] iR_{\mu \nu ab}\gamma ^{5}\Lambda _{\psi }\sigma ^{ab}\right) \left(
\Theta \left[ -f\right] I-\Theta \left[ f\right] i\gamma ^{5}\Lambda _{%
\overline{\psi }}\right) $

$=\Theta \left[ -f\right] R_{\mu \nu ab}\sigma ^{ab}+\Theta \left[ f\right]
R_{\mu \nu ab}\gamma ^{5}\Lambda _{\psi }\sigma ^{ab}\gamma ^{5}\Lambda _{%
\overline{\psi }}$

$=\Theta \left[ -f\right] R_{\mu \nu ab}\sigma ^{ab}+\Theta \left[ f\right]
R_{\mu \nu ab}\Lambda _{\psi }\sigma ^{ab}\Lambda _{\overline{\psi }}$

$=\Theta \left[ -f\right] R_{\mu \nu ab}\sigma ^{ab}+\Theta \left[ f\right]
R_{\mu \nu ab}\Lambda _{c}^{\;a}\Lambda _{d}^{\;b}\sigma ^{cd}$

$=\Theta \left[ -f\right] R_{\mu \nu \rho \sigma }e_{a}^{\;\rho
}e_{b}^{\;\sigma }\sigma ^{ab}+\Theta \left[ f\right] R_{\mu \nu \rho \sigma
}e_{a}^{\;\rho }e_{b}^{\;\sigma }\Lambda _{c}^{\;a}\Lambda _{d}^{\;b}\sigma
^{cd}$

$=R_{\mu \nu \rho \sigma }\left( \Theta \left[ -f\right] e_{a}^{\;\rho
}e_{b}^{\;\sigma }+\Theta \left[ f\right] \Lambda _{a}^{\;c}\Lambda
_{b}^{\;d}e_{c}^{\;\rho }e_{d}^{\;\sigma }\right) \sigma ^{ab}$.

From the gauge covariance of $\left[ D_{\mu },D_{\nu }\right] $ and $R$, we
can find a gauge covariant free-field term for $X_{\mu }$. \ We begin with
the expansion of $\left[ D_{\mu },D_{\nu }\right] $:%
\begin{eqnarray*}
\left[ D_{\mu },D_{\nu }\right] &=&\frac{\beta }{2}\left( \omega _{\mu ab}%
\left[ \sigma ^{ab},X_{\nu }\right] -\omega _{\nu ab}\left[ \sigma
^{ab},X_{\mu }\right] \right) \\
&&+\beta ^{2}\left[ X_{\mu },X_{\nu }\right] +\beta \left( \partial _{\mu
}X_{\nu }-\partial _{\nu }X_{\mu }\right) \\
&&+\left\{ \frac{1}{4}\omega _{\mu ab}\omega _{\nu cd}\left[ \sigma
^{ab},\sigma ^{cd}\right] +\frac{1}{2}\left( \partial _{\mu }\omega _{\nu
ab}-\partial _{\nu }\omega _{\mu ab}\right) \sigma ^{ab}\right\} \text{.}
\end{eqnarray*}%
The terms in $\left\{ ...\right\} $ give $R$ upon contraction with a couple
of vierbein. \ Because both $\left[ D_{\mu },D_{\nu }\right] $ and the $%
\left\{ ...\right\} $ terms transform gauge covariantly, we see that the
remaining terms - denoted as $H_{\mu \nu }$ - must also transform gauge
covariantly. \ We can now form a gauge covariant Lagrangian term from $%
H_{\mu \nu }$ as $Tr\left\{ H_{\mu \nu }H^{\mu \nu \dag }\right\} $ - this
is the previous free-field Lagrangian for $X_{\mu }$. \ The Hermitian action
we obtain from minimal coupling of $\omega _{\mu ab}$, $X_{\mu }$, and $\psi 
$ is thus [1]:

\begin{eqnarray*}
S &=&\dint \{\kappa R-m\overline{\psi }\psi +\frac{i}{2}e_{a}^{\;\mu }%
\overline{\psi }\gamma ^{a}\left( \partial _{\mu }\psi +\frac{1}{2}\omega
_{\mu bc}\sigma ^{bc}\psi +\beta X_{\mu }\psi \right) \}ed^{4}x \\
&&-\dint \left\{ \frac{i}{2}e_{a}^{\;\mu }\left( \partial _{\mu }\overline{%
\psi }-\frac{1}{2}\omega _{\mu bc}\overline{\psi }\sigma ^{bc}+\beta 
\overline{\psi }\gamma ^{0}X_{\mu }^{\dagger }\gamma ^{0}\right) \gamma
^{a}\psi \right\} ed^{4}x \\
&&+\dint \left\{ \frac{1}{4}Tr\left( H_{\mu \nu }H^{\mu \nu \dag }\right)
\right\} ed^{4}x\text{, where}
\end{eqnarray*}%
$e=\det \left( e_{a}^{\;\mu }\right) $ and $H_{\mu \nu }=\frac{\beta }{2}%
\left( \omega _{\mu ab}\left[ \sigma ^{ab},X_{\nu }\right] -\omega _{\nu ab}%
\left[ \sigma ^{ab},X_{\mu }\right] \right) +\beta ^{2}\left[ X_{\mu
},X_{\nu }\right] +\beta \left( \partial _{\mu }X_{\nu }-\partial _{\nu
}X_{\mu }\right) $. \ The resulting equations of motion are found in [1].

The $x_{\mu ab}$ terms are the obvious terms to examine as possible
explanations for current gravitational issues because they couple to the
matter fields in the same manner as the spin connection. \ Experimental
speculations and crude predictions can be found in [1, 2]. \ Also, the $%
x_{\mu ab}$ field equations should explain the Baryonic Tully-Fisher law and
the Faber-Jackson law. \ However, difficulties encountered by the author in
trying to explain the slope appearing in the Tully-Fisher law using a gauge
theory of the CPT symmetry - or \textit{any }force obeying an inverse square
law - have led to consideration of a slightly different free-field
Lagrangian of the gauge theory of the CPT symmetry. \ The physical
implications of the "chiral" terms, $x_{\mu I}$ and $x_{\mu 5}$, and their
associated anomalies are not developed enough to warrant a discussion.
\bigskip

THE NEW LAGRANGIAN
\bigskip

To proceed further, we recall that the CPT symmetry is born from the union
of special relativity with quantum theory; therefore, we intuitively expect $%
X_{\mu }$ to be a "bridge" connecting the two [1,2]. \ The free-field
Lagrangians of the quantum field theories appearing in the Standard Model
are based on (using the applicable covariant derivatives) $Tr\left\{ \left[
D_{\mu },D_{\nu }\right] \left[ D^{\mu },D^{\nu }\right] ^{\dag }\right\} $,
and the free-field Lagrangian $R$ of general relativity is of the form
(obtained from $\omega _{\mu ab}$) $\left[ D_{\mu },D_{\nu }\right] $
appropriately contracted. \ Therefore, we expect $X_{\mu }$ to appear in 
\textit{both} types of free-field Lagrangians. \ The previous free-field
Lagrangian did not satisfy this criterion.

A simple, intuitive candidate is to retain the Lagrangian discussed above,
but replace the Einstein-Hilbert $R$ by a similar term formed by
substituting $\frac{1}{2}\omega _{\mu ab}$ with $\frac{1}{2}\omega _{\mu
ab}+\beta x_{\mu ab}$ instead. \ We denote this modified curvature scalar by 
$R_{X\omega }$. \ This new free-field Lagrangian formed from $Tr\left\{
H_{\mu \nu }H^{\mu \nu \dag }\right\} $ and $R_{X\omega }$ satisfies all the
criteria we require - provided that the new candidate transforms gauge
covariantly, too. \ We now turn our attention to this task. \ We begin by
again expanding $\left[ D_{\mu },D_{\nu }\right] $:%
\begin{eqnarray*}
\left[ D_{\mu },D_{\nu }\right] &=&\frac{\beta }{2}\left( \omega _{\mu ab}%
\left[ \sigma ^{ab},X_{\nu }\right] -\omega _{\nu ab}\left[ \sigma
^{ab},X_{\mu }\right] \right) \\
&&+\beta ^{2}\left[ X_{\mu },X_{\nu }\right] +\beta \left( \partial _{\mu
}X_{\nu }-\partial _{\nu }X_{\mu }\right) \\
&&+\left\{ \frac{1}{4}\omega _{\mu ab}\omega _{\nu cd}\left[ \sigma
^{ab},\sigma ^{cd}\right] +\frac{1}{2}\left( \partial _{\mu }\omega _{\nu
ab}-\partial _{\nu }\omega _{\mu ab}\right) \sigma ^{ab}\right\} \text{.}
\end{eqnarray*}
The term $R_{X\omega }$ is formed from the contraction (with a couple of
vierbein) of:

\[
\left[ D_{\mu },D_{\nu }\right] -\beta \left\{ \left( \partial _{\mu }x_{\nu
I}-\partial _{\nu }x_{\mu I}\right) I+\left( \partial _{\mu }x_{\nu
5}-\partial _{\nu }x_{\mu 5}\right) \gamma ^{5}\right\} . 
\]%
\ So, if we can show that $\beta \left\{ \left( \partial _{\mu }x_{\nu
I}-\partial _{\nu }x_{\mu I}\right) I+\left( \partial _{\mu }x_{\nu
5}-\partial _{\nu }x_{\mu 5}\right) \gamma ^{5}\right\} $ transforms gauge
covariantly, then so does $R_{X\omega }$.

We begin by defining $F_{\mu \nu }\equiv \left[ D_{\mu },D_{\nu }\right]
=\sum\limits_{i}y_{i}\Gamma ^{i}$, where the $\Gamma ^{i}$ are the eight
linearly independent matrices $I$, $\gamma ^{5}$, and $\sigma ^{ab}$; and
the $y_{i}$ are their coefficients (we will not be using the Einstein
summation convention for this proof). \ We define the transformation of $%
F_{\mu \nu }$ under gauge $CPT\Lambda $ (i.e., $U$ from above) as $F_{\mu
\nu }^{^{\prime }}=\sum\limits_{i}y_{i}^{^{\prime }}\Gamma ^{i}$. \ We also
know that $F_{\mu \nu }^{^{\prime }}=UF_{\mu \nu }U^{-1}$, therefore, $%
\sum\limits_{i}y_{i}^{^{\prime }}\Gamma ^{i}=\sum\limits_{i}y_{i}U\Gamma
^{i}U^{-1}$. \ Let $U\Gamma ^{i}U^{-1}=\sum\limits_{j}a_{j}^{i}\Gamma ^{j}$,
then $\sum\limits_{i}y_{i}^{^{\prime }}\Gamma
^{i}=\sum\limits_{i}\sum\limits_{j}y_{i}a_{j}^{i}\Gamma
^{j}=\sum\limits_{i}\sum\limits_{j}y_{j}a_{i}^{j}\Gamma ^{i}$, so $%
y_{i}^{^{\prime }}=\sum\limits_{j}y_{j}a_{i}^{j}$. \ We note that $%
UIU^{-1}=I $, $U\gamma ^{5}U^{-1}=\gamma ^{5}$, and $U\sigma
^{ab}U^{-1}=\sum\limits_{cd}r_{cd}\sigma ^{cd}$, where the $r_{cd}$ are just
the coefficients appearing in the transformation of $\sigma ^{ab}$. \
Therefore, the transformation of the terms containing $I$ or $\gamma ^{5}$
are not mixed up with the $\sigma ^{ab}$ terms. \ We now want to know what
is required to have a single $y_{i}\Gamma ^{i}$ term transform gauge
covariantly. \ In other words, what is required to have $y_{i}^{^{\prime
}}\Gamma ^{i}=y_{i}U\Gamma ^{i}U^{-1}$? \ Or, equivalently, what is required
for $\sum\limits_{j}y_{j}a_{i}^{j}\Gamma ^{i}=y_{i}U\Gamma ^{i}U^{-1}$? \
Hitting both sides of this equation by $\left( \Gamma ^{i}\right) ^{-1}$
gives $\sum\limits_{j}y_{j}a_{i}^{j}I=y_{i}\left( \Gamma ^{i}\right)
^{-1}U\Gamma ^{i}U^{-1}$. \ So, for this equation to be true, the following
two conditions must be met: \ $I=\left( \Gamma ^{i}\right) ^{-1}U\Gamma
^{i}U^{-1}$, and $\sum\limits_{j}y_{j}a_{i}^{j}=y_{i}$. \ The first
condition is met for $\Gamma ^{i}=I$ and $\Gamma ^{i}=\gamma ^{5}$. \ The
second condition is met for the $U$ of gauge $CPT\Lambda $ because $%
a_{i}^{j}=\delta _{i}^{j}$ for $\Gamma ^{i}=I$ and $\Gamma ^{i}=\gamma ^{5}$%
. \ Therefore, the entire term $\beta \left\{ \left( \partial _{\mu }x_{\nu
I}-\partial _{\nu }x_{\mu I}\right) I+\left( \partial _{\mu }x_{\nu
5}-\partial _{\nu }x_{\mu 5}\right) \gamma ^{5}\right\} $ transforms gauge
covariantly - and so does $R_{X\omega }$. \ Therefore, without consideration
of the appearance of the chiral anomaly or any other anomalies which might
appear, the Hermitian action we obtain from minimal coupling of $\omega
_{\mu ab}$, $X_{\mu }$, and $\psi $ is thus:

\begin{eqnarray*}
S &=&\dint \{\kappa R_{X\omega }-m\overline{\psi }\psi +\frac{i}{2}%
e_{a}^{\;\mu }\overline{\psi }\gamma ^{a}\left( \partial _{\mu }\psi +\frac{1%
}{2}\omega _{\mu bc}\sigma ^{bc}\psi +\beta X_{\mu }\psi \right) \}ed^{4}x \\
&&-\dint \left\{ \frac{i}{2}e_{a}^{\;\mu }\left( \partial _{\mu }\overline{%
\psi }-\frac{1}{2}\omega _{\mu bc}\overline{\psi }\sigma ^{bc}+\beta 
\overline{\psi }\gamma ^{0}X_{\mu }^{\dagger }\gamma ^{0}\right) \gamma
^{a}\psi \right\} ed^{4}x \\
&&+\dint \left\{ \frac{1}{4}Tr\left( H_{\mu \nu }H^{\mu \nu \dag }\right)
\right\} ed^{4}x\text{.}
\end{eqnarray*}

The appearance of $X_{\mu }$ in both $R_{X\omega }$ and $H_{\mu \nu }$ leads
to interesting equations of motion for the $x_{\mu ab}$ terms. \ Also, the
replacement of $R$ by $R_{X\omega }$ provides the intuition that the
gravitational lensing effects attributed to the hypothetical dark matter are
explained by gauge $CPT\Lambda $ instead. \ Finally, we note that in the
limit as $X_{\mu }\rightarrow 0$, we obviously recover general relativity.
\bigskip

This paper is dedicated to the memory of the author's Mother.
\bigskip

REFERENCES
\bigskip

[1] K. Koltko, "Gauge CPT as a Possible Alternative to the Dark Matter
Hypothesis," arXiv: \ 1308.6341 [physics.gen-ph] and cross-listed as
[astro-ph.GA] (2013).

[2] K. Koltko, \textit{Attempts to Find Additional Dynamical Degrees of
Freedom in Spacetime Using Topological and Geometric Methods,} 2000 Ph.D.
Thesis, University of South Carolina.

\end{document}